\documentclass[twocolumn,showpacs,preprintnumbers,amsmath,amssymb,showkeys]{revtex4}

\usepackage{graphicx}
\usepackage{dcolumn}
\usepackage{bm}
\usepackage{color}

\begin{document}

\preprint{Superlattices and Microstructures, Volume 25, Issues 1-2, January 1999, Pages 431-438}

\title{Optimized minigaps for negative differential resistance creation in strongly $\delta$-doped (1D) superlattices}

\author{T. Ferrus \footnote{Present address : Hitachi Cambridge Laboratory, J. J. Thomson Avenue, CB3 0HE, Cambridge, United Kingdom}}
\email{taf25@cam.ac.uk}
\author{B. Goutiers}
\author{L. Ressier}
\author{J. P. Peyrade}
\affiliation {Institut National des Sciences appliqu\'ees, 135 Avenue de Rangueil, 31077 Toulouse, France}
\author{J. A. Porto}
\author{J. S\'anchez-dehesa}
\affiliation{Departamento de F\'isica Te\'orica de la Materia Condensada, Universidad Aut\'onoma de Madrid, Madrid, 28049, Spain}


\date{\today}

\begin{abstract}

The \textit{atomic saw method} uses the passage of dislocations in two-dimensional (2D) quantum-well superlattices to create periodic slipping layers and one-dimensional (1D) quantum wire superlattices. The effects of this space structuring of the samples on the allowed energies are analysed in the case of GaAs d-doped superlattices. If they are sufficiently large, the various minigaps appearing in the 1D band structure could be responsible for the presence of negative differential resistance (NDR) with high critical current in these systems. The purpose is to determine the evolution of the minigaps in terms of the sample parameters and to obtain the means to determine both the 2D and 1D structural characteristics where NDR could appear.

\end{abstract}

\maketitle

Full article available at doi:10.1006/spmi.1998.0670

see also erratum at doi:10.1006/spmi.1998.0702

\end{document}